%% file: HIN-19-005_temp.tex
\pdfoutput=1
\documentclass[11pt,twoside,a4paper,cmspaper,final,collab]{cms-tdr}
\def\svnVersion{7637ccc}\def\svnDate{2022/01/06}\def\cmsCernNoTag{CERN-EP-2021-023}\def\cmsCernDate{\today}\def\cmsMessage{Published in Physical Review Letters as \href{http://dx.doi.org/10.1103/PhysRevLett.128.032001}{\doi{10.1103/PhysRevLett.128.032001}.}}
\begin{document}\cmsNoteHeader{HIN-19-005}

\newcommand{\Jpsi}{\PJGy}
\newcommand{\xthree}{\ensuremath{\PX(3872)}\xspace}
\newcommand{\psitwoS}{\ensuremath{\Pgy}\xspace}
\newcommand{\xandpsi}{\xthree and \psitwoS}
\newcommand{\Jpsipipi}{\ensuremath{\PJGy\,\Pgpp\Pgpm}\xspace}

\newcommand{\pythia}{\PYTHIA}
\newcommand{\evtgen}{\EVTGEN}
\newcommand{\hydjet}{\HYDJET}
\newcommand{\photos}{\PHOTOS}
\newcommand{\pp}{\ensuremath{\Pp\Pp}\xspace}
\newcommand{\PbPb}{\ensuremath{\mathrm{Pb}\mathrm{Pb}}\xspace}

\newcommand{\sqrts}{\ensuremath{\sqrt{s}}}
\newcommand{\raa}{\ensuremath{R_{\mathrm{AA}}}\xspace}
\newcommand{\fprompt}{\ensuremath{f_{\text{prompt}}}\xspace}
\newcommand{\chisq}{\ensuremath{\chi^{2}}\xspace}
\newcommand{\ereco}{\ensuremath{\epsilon_{\text{reco}}}}
\newcommand{\esel}{\ensuremath{\epsilon_{\text{sel}}}}
\newcommand{\lxy}{\ensuremath{l_{xy}}\xspace}
\newcommand{\Nincl}{\ensuremath{N_{\text{incl}}}}
\newcommand{\Benr}{\ensuremath{\Pb\text{\mbox{-}enr}}\xspace}
\newcommand{\NBenr}{\ensuremath{N_{\Benr}}\xspace}

\newcommand{\figref}[1]{Fig.~\ref{#1}}
\newcommand{\eqqref}[1]{Eq.~\eqref{#1}}

\cmsNoteHeader{HIN-19-005} 
\title{Evidence for X(3872) in \texorpdfstring{\PbPb}{PbPb} collisions and studies of its prompt production at \texorpdfstring{$\sqrtsNN= 5.02\TeV$}{sqrt(sNN) = 5.02 TeV}}

\date{\today}

\abstract{
The first evidence for \xthree production in relativistic heavy ion collisions is reported.
The \xthree production is studied in lead-lead (\PbPb) collisions at a center-of-mass energy of $\sqrtsNN = 5.02\TeV$ per nucleon pair, using the decay chain $\xthree\to \PJGy\,\Pgpp\Pgpm\to \mu^+\mu^-\Pgpp\Pgpm$. The data were recorded with the CMS detector in 2018 and correspond to an integrated luminosity of 1.7\nbinv. The measurement is performed in the rapidity and transverse momentum ranges $\abs{y}<1.6$ and $15<\pt < 50\GeVc$. The significance of the inclusive \xthree signal is 4.2 standard deviations. The prompt \xthree to \psitwoS yield ratio is found to be $\rho^{\PbPb} = 1.08\pm0.49\stat\pm0.52\syst$, to be compared with typical values of 0.1 for \pp collisions. This result provides a unique experimental input to theoretical models of the \xthree production mechanism, and of the nature of this exotic state.}

\hypersetup{
pdfauthor={CMS Collaboration},
pdftitle={Evidence for X(3872) in PbPb collisions and studies of its prompt production at sqrt(sNN) = 5.02 TeV},
pdfsubject={CMS},
pdfkeywords={CMS, quark-gluon plasma, heavy ions, X(3872)}}

\maketitle 

The \xthree, also known as \PGccDoP{3872}, is an exotic particle that was first observed by the Belle Collaboration~\cite{Choi:2003ue}, and then confirmed and studied by other experiments at electron-positron~\cite{Aubert:2004ns,Ablikim:2013dyn} and hadron colliders~\cite{Acosta:2003zx,Abazov:2004kp,Aaij:2011sn,Chatrchyan:2013cld,Aaboud:2016vzw}. The quantum numbers of the \xthree have been narrowed down by the CDF~\cite{Abulencia:2006ma}, and later determined to be $J^{PC}=1^{++}$ by the LHCb~\cite{Aaij:2013zoa} Collaborations. However, the nature of this particle is still not fully understood and interpretations in terms of conventional charmonium (a bound state of charm-anticharm quarks), \PDstz\PaDz molecules~\cite{Tornqvist:2004qy}, tetraquark states~\cite{Maiani:2004vq}, and their admixture~\cite{Hanhart:2011jz} have been proposed. The production and survival of the \xthree in a quark-gluon plasma (QGP), a deconfined state of quarks and gluons~\cite{Collins:1974ky,Shuryak:1978ij}, or after the QGP, in a hadronic phase, is expected to depend upon the \xthree's internal structure~\cite{Zhang:2020dwn,Wu:2020zbx}. Thus, the recent large data set of lead-lead (PbPb) collisions at a center-of-mass energy of $\sqrtsNN = 5.02\TeV$ per nucleon pair, delivered by the Large Hadron Collider (LHC) at CERN at the end of 2018, opened new opportunities to probe the nature of this exotic state~\cite{Cho:2010db,Cho:2011ew,Torres:2014fxa}. 

It is expected that in relativistic heavy ion collisions, the formation of the QGP could enhance or suppress the production of the \xthree particle. Coalescence mechanisms could enhance the \xthree production yield~\cite{Zhang:2020dwn,Cho:2011ew}. These mechanisms can be modeled via the overlap of the density matrix of the constituents in an emission source with the Wigner function of the produced particle~\cite{SATO1981153}. Therefore, the enhancement of the \xthree production in the QGP would depend on the spatial configuration of the exotic state. Moreover, a longer distance between the quarks and antiquarks that constitute the state could also lead to a higher \xthree dissociation rate, similar to that from the mechanism of quarkonium suppression in heavy ion collisions~\cite{Matsui:1986dk}. Therefore, the study of the \xthree state in the QGP may be used as a tool to distinguish a compact tetraquark configuration with a radius $\sim$0.3\unit{fm} from a molecular state with a radius greater than 1.5\unit{fm}~\cite{Abreu:2016qci}. Such a measurement would be complementary to the recent evidence for the radiative decay $\xthree\to\Pgy\gamma$ in proton-proton (\pp) collisions reported by LHCb Collaboration~\cite{Aaij:2014ala}, which does not support a pure \PDstz\PaD molecular interpretation. In addition, measurements of prompt \xthree production could provide an interesting test of the statistical hadronization model, which assumes that the produced matter is in thermodynamic equilibrium at the phase transition to hadrons~\cite{Andronic:2005yp,Andronic:2017pug}. 

In this Letter, the first evidence for \xthree production in \PbPb collisions at $\sqrtsNN= 5.02\TeV$ is reported. The \PbPb sample corresponds to an integrated luminosity of 1.7\nbinv. The \xthree candidates are reconstructed through the decay chain \mbox{$\xthree\to \PJGy\,\Pgpp\Pgpm\to \mu^+\mu^-\Pgpp\Pgpm$}, and are measured in the $15<\pt<50$\GeVc and $\abs{y}<1.6$ kinematic region. At the LHC energies, the inclusive \xthree yields in \pp and \PbPb collisions contain a significant nonprompt contribution coming from \Pb hadron decays~\cite{Aaboud:2016vzw}. The nonprompt \xthree component is related to the medium-modified beauty hadron production in heavy ion collisions, which is out of the scope of this paper. Here, we focus on the prompt component, from charm quark fragmentation, for which the ratio $\rho^{i}$ ($i$ is \pp or \PbPb) between the corrected yields of \xthree and \psitwoS mesons (where the \psitwoS is reconstructed with the same final-state particles in order to reduce  systematic uncertainties) is presented: 
\begin{linenomath} 
\begin{equation}
  \rho^{i} = \frac{N^{\xthree\to\PJGy\,\pi\pi}_{i}}{N^{\psitwoS\to\PJGy\,\pi\pi}_{i}}.
  \label{eq:result.r}
\end{equation} 
 \end{linenomath}
 The ratios in \pp and \PbPb collisions are connected to the nuclear modification factors $\raa^{\xthree}$ and $\raa^{\psitwoS}$ (the meson yield ratio in nucleus-nucleus and \pp interactions normalized by the number of inelastic nucleon-nucleon collisions) via the following relation: 
\begin{linenomath} 
\begin{equation}     
   \rho^{\PbPb} = \rho^{\pp} \frac{\raa^{\xthree}}{\raa^{\psitwoS}}.
\label{Eq:raa}
\end{equation}
 \end{linenomath}
 
The CMS apparatus~\cite{CMS:2008xjf} is a multipurpose, nearly hermetic detector, designed to trigger on~\cite{CMS:2020cmk,CMS:2016ngn} and identify electrons, muons, photons, and (charged and neutral) hadrons~\cite{CMS:2015xaf,CMS:2018rym,CMS:2015myp,CMS:2014pgm}. A global reconstruction algorithm~\cite{CMS:2017yfk} combines the information provided by the all-silicon inner tracker and by the crystal electromagnetic and brass-scintillator hadron calorimeters, operating inside a 3.8\unit{T} superconducting solenoid, with data from gas-ionization muon detectors interleaved with the solenoid return yoke. Information from the hadron forward (HF) calorimeter is used for performing offline event selection and determining centrality (the degree of overlap between the two colliding nuclei). The results refer to the collisions with 0--90\% centrality, i.e., the top 90\% events based on the total transverse energy deposition in both HF detectors~\cite{Chatrchyan:2011sx}, which corresponds to the 90\% of collisions having the largest overlap of the two nuclei.

Events of interest were selected in real-time using the CMS two-tiered trigger system: the first level (L1), composed of custom hardware processors~\cite{CMS:2020cmk}, and the high-level trigger (HLT), consisting of a farm of processors running a version of the full event reconstruction software optimized for fast processing~\cite{CMS:2016ngn}. The selection required the presence of two muon candidates, with at least one muon reconstructed in the outer muon spectrometer, and one muon reconstructed using information from both the outer muon spectrometer and the inner tracker. The dimuon candidate invariant mass is required to be \mbox{$1 < m_{\mu\mu} < 5\GeVcc$}. For the offline analysis, events have to pass a set of selection criteria designed to reject events from background processes (beam-gas collisions, beam scraping events and electromagnetic interactions) as described in Ref.~\cite{Khachatryan:2016odn}. Events are required to have at least one reconstructed primary interaction vertex formed by two or more tracks, with a distance from the center of the nominal interaction point of less than 15\cm along the beam axis. The shapes of the clusters in the pixel detector have to be compatible with those expected from particles produced by a \PbPb collision~\cite{Khachatryan:2010xs}. In order to select hadronic collisions, the \PbPb events are also required to have at least two towers (\ie, a geometrically defined group of calorimeter cells) in each of the HF detectors with total energy deposits of more than 4\GeV per tower. This analysis is restricted to events within centrality 0--90\%, for which the hadronic event selection is fully efficient. Multiple-collision events (pileup) have a negligible effect on the measurement, since the average number of additional collisions per bunch crossing is approximately 0.002.

Dedicated \PbPb \xthree and \psitwoS Monte Carlo (MC) simulated samples were generated in order to estimate the acceptance and selection efficiencies, 
to study the background components, and to evaluate systematic uncertainties. 
The \pythia~8 v212~\cite{Sjostrand:2014zea} Tune CP5~\cite{Sirunyan:2019dfx} was used to generate the \xthree and \psitwoS signals at $\sqrtsNN=5.02\TeV$. It was assumed that \xandpsi are unpolarized.
Since the \xthree cannot be generated by \pythia, the $\PGccDoP{1P}$ particle is used, with a modified mass of 3.8716\GeVcc~\cite{PDG2020}. The $\PGccDoP{1P}$ has the quantum numbers $J^{PC} = 1^{++}$, identical to those of the \xthree.
The \xthree particle was forced to decay into \Jpsipipi (assuming the $\rho$ resonance dominates the pion pair spectrum~\cite{Chatrchyan:2013cld,Aaboud:2016vzw}), followed by the \Jpsi meson decaying into two muons.
Final-state radiation was generated using \photos~2.0~\cite{Barberio:1990ms}.  The $\PGccDoP{1P} \to \Jpsipipi$ decay is generated with \evtgen. The samples with prompt (fragmenting in charm quarks) and nonprompt (originating from \Pb hadron decays) \psitwoS/\xthree production are generated separately. Each \pythia event is embedded in a \PbPb collision event generated with \HYDJET~1.8~\cite{Lokhtin:2005px}, which is tuned to reproduce global event properties such as the charged-hadron \pt spectrum and particle multiplicity.

{\tolerance=800
The \xthree signal is extracted in the following steps. Each muon candidate must be matched to a triggered muon and have $\pt^{\mu} > 3.5\GeVc$ in the interval $\abs{\eta^{\mu}}<1.2$, $\pt^{\mu} > (5.47\text{--}1.89\,\abs{\eta^{\mu}})\GeVc$ in the interval $1.2<\abs{\eta^{\mu}}<2.1$, or $\pt^{\mu} > 1.5\GeVc$ in the forward region $2.1 < \abs{\eta^{\mu}} < 2.4$. Two muons of opposite sign, with an invariant mass within $\pm$150\MeVcc of the world-average $\PJGy$ meson mass~\cite{PDG2020} are selected to reconstruct a $\PJGy$ candidate. The opposite-sign muon pairs are fitted with a common vertex constraint and are kept if the \chisq probability of the fit is greater than 1\%, thus reducing the background from charm and beauty hadron semileptonic decays. The \xthree and \psitwoS candidates are built by combining the \Jpsi candidates with two additional tracks, which have $ \pt > 0.9\GeVc$, $\abs{\eta}<2.4$ and pass a high purity selection, assumed to be produced by two pions. Only candidates that have $15<\pt<50$\GeVc and $\abs{y}<1.6$ are considered. Then, a kinematic fit to the \Jpsipipi system is performed, requiring that the four tracks originate from a common vertex and forcing the mass of the dimuon pair to be equal to the nominal \Jpsi mass~\cite{PDG2020}. The selection is further optimized, using a boosted decision tree (BDT) algorithm~\cite{Hocker:2007ht}. The \xthree decay vertex probability, the radial distance between the pion and the \Jpsi candidate momentum vectors ($\sqrt{\smash[b]{(\Delta\eta)^2+(\Delta\phi)^2}}$, where $\phi$ is the azimuthal angle), and the \pt of each pion are used in the BDT algorithm to distinguish signal and combinatorial background formed by random combinations of tracks. For the multivariate training, the \xthree signal sample is taken from simulation, and the background sample consists of data from the sidebands of the \xthree meson peak (\ie, $0.07 < \abs{m - m_{\xthree}} < 0.128$\GeVcc). The nominal BDT selection is chosen in order to maximize the statistical significance of \xthree, defined as $S/\sqrt{(S+B)}$ where $S$ and $B$ are the numbers of \xthree and background in the signal region, respectively. Because there is no reliable theoretical calculation available for the \xthree production, an estimation of the \xthree cross-section is obtained from data, by extracting the signal yield in data when applying a tight BDT cut. This is used in conjunction with the MC-calculated efficiency to obtain the yield for each BDT cutoff value. The BDT selection determined in this way is applied to entries in the whole invariant mass range from 3.62 to 4\GeVcc in data.
\par}

\begin{figure}[bhpt]
\centering
\includegraphics[width=0.45\textwidth]{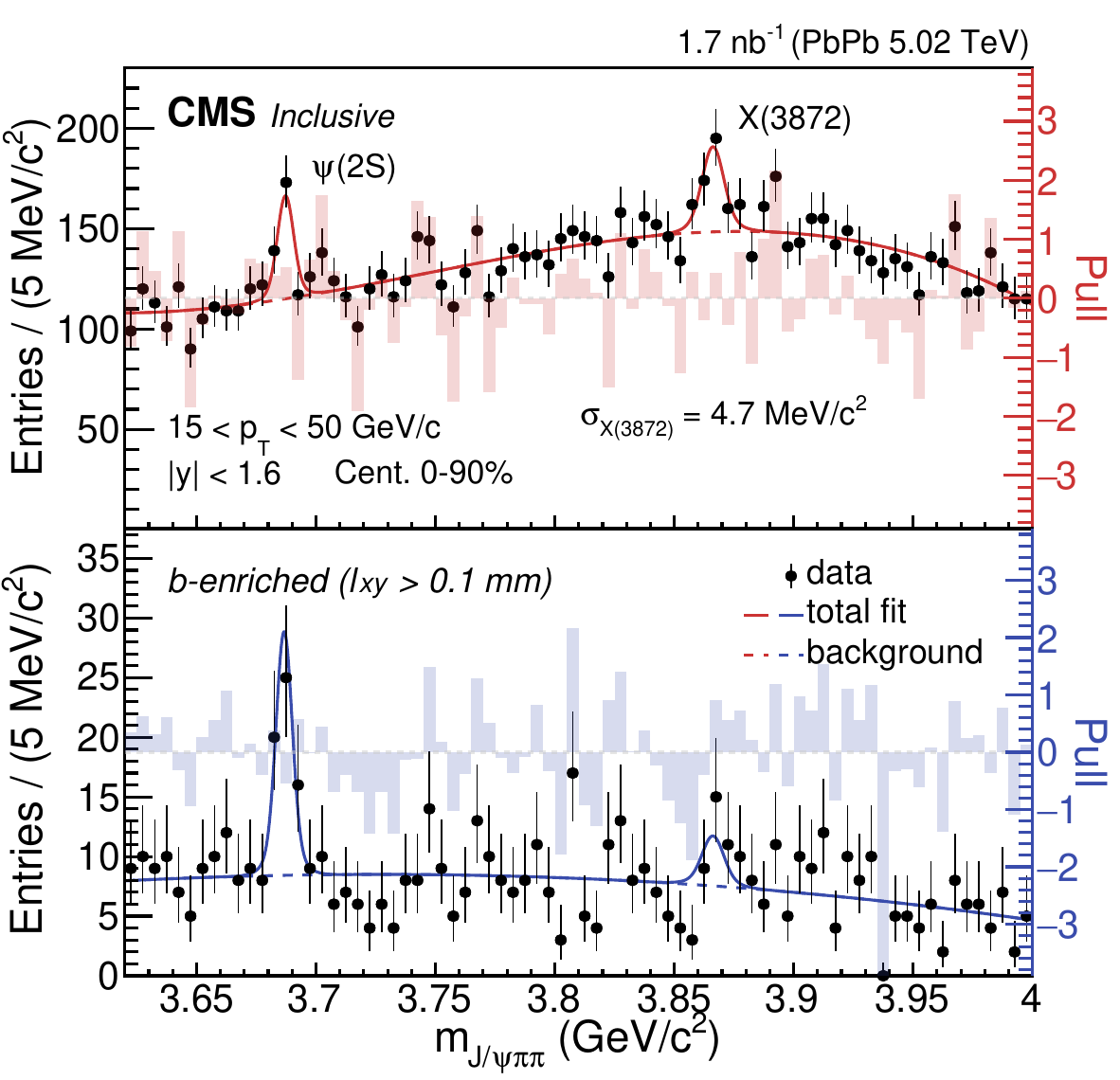}
\caption{Invariant mass distribution of $m_{\mu\mu\pi\pi}$ in \PbPb collisions, for the inclusive (upper) and \Pb-enriched (bottom) samples. The vertical lines represent statistical uncertainties in the data. The results of the unbinned maximum-likelihood fits for the signal$+$background and background alone are also shown by the solid and dashed lines, respectively. The pull distribution is represented by the shaded bars. The \xthree peak mass resolution, $\sigma_{\xthree}$, calculated  at the half-maximum of the signal-shape distribution, is also listed for reference.}
\label{fig:invmass}
\end{figure}

The raw inclusive yields of \xandpsi are extracted by an extended unbinned maximum-likelihood fit. A double-Gaussian function with a common mean but independent widths is used to model the signal component for each of the \xandpsi peaks. This was preferred to a single-Gaussian or a Breit--Wigner function since it described better (\ie, superior \chisq of the fit) the signal shape in MC simulations. For describing the combinatorial background, mostly produced by the random combination of a \Jpsi candidate with tracks that are not coming from \xthree or \psitwoS decay, a 4th-order polynomial is used, which gives the best fit in terms of \chisq per degrees of freedom and stability during all studies. For the signal, only the magnitude of the two peaks is left free in the fit, the rest (the mean and widths of the two Gaussian functions, as well as their relative contribution to the signal yield in either \xthree or \psitwoS peaks) are set to the values derived from simulation. The five parameters of the combinatorial background are all allowed to float. The invariant mass range considered for the fit is 3.62 to 4\GeVcc. The invariant mass fits for both the inclusive and nonprompt samples, with BDT selection optimized for \xthree, are shown in \figref{fig:invmass}. The significance of the inclusive \xthree signal against background-only hypothesis is 4.2 standard deviations. The systematic uncertainty (described below) contributing to this significance is the one related to the \xthree invariant mass fit. After performing a likelihood scan for each alternative signal and background shape considered, the significance was calculated as the square root of the logarithm of the profile likelihood ratio where the signal is zero, with the smallest value obtained being chosen among all scans.

{\tolerance=800
The contribution from \Pb hadron decays is subtracted from the inclusive result using the ``pseudo-proper'' decay length \lxy, defined as the distance in the transverse plane, $L_{xy}$, between the vertex formed by the 4-tracks and the primary vertex, corrected by the transverse Lorentz boost of the candidate: \mbox{$\lxy = L_{xy}\,m_{\Jpsi\pi\pi}c/\abs{\ptvec}$}.
The prompt-component fraction (\fprompt) is estimated using a cutoff-based method in the following way. Since the \lxy of the prompt component was found in MC studies to be smaller than 0.1\mm, the prompt fraction, \fprompt, can be derived from: i) the raw inclusive  yield, \Nincl, obtained from the fit to the invariant mass distributions of all candidates, shown in \figref{fig:invmass}, and ii) \NBenr, the ``\Pb-enriched'' yield, obtained from a fit to the invariant mass distribution only containing candidates that passed the selection $\lxy>0.1\mm$, also shown in \figref{fig:invmass}. In addition, \NBenr has to be corrected to account for nonprompt candidates, with $\lxy<0.1\mm$, that have been missed:  $f_\text{nonprompt}^{\Benr} =N_\text{nonprompt}(\lxy>0.1\mm)/N_\text{nonprompt}$. The correction was obtained from simulation. The raw prompt fraction is
then calculated as:
\begin{linenomath} 
\begin{equation}     
   \fprompt = 1 - \frac{\NBenr / f^{\Benr}_{\text{nonprompt}}}{\Nincl},
\end{equation}
 \end{linenomath}
separately for \xthree and \psitwoS states.
The corrected yield of the prompt component can then be derived as:
\begin{linenomath} 
\begin{equation}
  N^{i\to\PJGy\pi\pi} = N_{\text{raw}}^{i} \, \fprompt^{i} / (\alpha\,\ereco\,\esel)^{i},
\label{eq:result.N}
\end{equation}
 \end{linenomath}
where $i$ is \xthree or \psitwoS, $\alpha$ is the acceptance, \ereco\ is the candidate reconstruction efficiency and \esel\ is the candidate selection efficiency. Since the two states are reconstructed in the same decay channel and are relatively close in mass, their corresponding $\alpha\,\ereco$ values are similar. The choice of the BDT optimization criteria results in \esel\ being higher for the \xthree than for the \psitwoS.
\par}

The measurement of $\rho^{\PbPb}$ is affected by several sources of systematic uncertainty, arising from the candidate selection, invariant mass fit, and efficiency corrections. To estimate the systematic uncertainty associated with the BDT selection, the BDT cutoff values are varied within a range that allows a robust invariant mass fit procedure (\ie, signal statistical significance larger than 2), and for each variation all factors in Eq.~\eqref{eq:result.N} are recalculated, separately for \xandpsi. The maximum difference of the final $\rho^{\PbPb}$ value from the nominal result (40\%) is quoted as the systematic uncertainty. The relatively large $\rho^{\PbPb}$ uncertainty associated with BDT cutoff value is the convolution of mainly two causes: the BDT variables distribution differences in data and MC samples for the \xthree meson, and the statistical limitation of the signal in data. The largest differences ($\sim$2 standard deviations) between data and MC samples are in the distributions of the \pt of the pions, and the radial distance between the pion and the \Jpsi candidate momentum vectors.

The uncertainty in the invariant mass fit (8.0\%) is calculated by adding in quadrature the maximum deviations from the nominal result to that found using two alternative fitting functions for both signal and background. For the signal, one variation consists of using a triple-Gaussian function, while for the other the signal width of the nominal fit is allowed to float to account for the resolution difference between data and MC. Other choices for the signal shape (\eg, one-Gaussian function) were not considered because of their poor-quality fits. For the background, the fit function is changed once to a third-order polynomial (as an exponential function or lower-order polynomials could not describe the data), and the fit range is also changed from 3.62--4\GeVcc to 3.62--3.9\GeVcc to exclude the right-hand shoulder.

The efficiency corrections obtained from simulation are sensitive to how well the \pt spectrum of the \xandpsi candidates is  modeled. The uncertainty related to the simulated \pt shape is evaluated by comparing the reconstruction and selection efficiencies calculated using the default \pythia MC sample, with another MC sample in which the \pt distributions of \xandpsi are tuned to reproduce the extracted \xandpsi \pt and $y$ spectra obtained in data, by performing mass fits in bins of \xandpsi \pt and $y$. The \pt and $y$ spectra of the alternative MC samples are allowed to vary within the statistical uncertainties in data. The mean of the differences between efficiencies from the alternative MC samples and the default \pythia MC due to the variation of \pt and $y$ spectra, which is 13\%, is quoted as the systematic uncertainty.

The uncertainties in the trigger efficiency, in the muon reconstruction and identification are evaluated using single muons from \PJGy meson decays in both simulated and collision data, with the tag-and-probe method~\cite{Khachatryan:2010xn,Chatrchyan:2012xi}. This combined uncertainty is found to be negligible, below 1\%.
Scale factors, calculated as the ratio of data to simulated efficiencies as a function of $\pt^\mu$ and $\eta^\mu$, are applied to each dimuon pair on a muon-by-muon basis. The uncertainties of the scaling factors from tag-and-probe studies are quoted as systematic uncertainties. 

To estimate the uncertainty in the prompt fraction arising from potential differences between the resolution in data and simulation, a template fit of the \lxy distribution in data is performed using prompt and nonprompt \lxy templates from simulation. Data are binned in \lxy, and an invariant mass fit is performed to extract the inclusive yield in each \lxy bin. This background-subtracted \lxy distribution is then fitted using a two component fit, which includes the prompt and nonprompt \lxy templates from simulation. The widths of the simulated DCA distributions are varied by a floating scale factor, and the best simulated smearing scale factor to match data is determined by minimizing the \chisq of the two-component fit. The difference between the ratio of the prompt fractions of \xthree to \psitwoS using the template fit method and the nominal result (8.1\%) is quoted as a systematic uncertainty.

When calculating the uncertainties in the ratio of the acceptance-corrected yields of prompt \xthree production over prompt \psitwoS production, the uncertainties of \xthree and \psitwoS yields are assumed to be independent except for the systematic uncertainties from muon reconstruction, efficiencies, and prompt fractions. 

\begin{figure}[bhpt]
\centering
\includegraphics[width=0.45\textwidth]{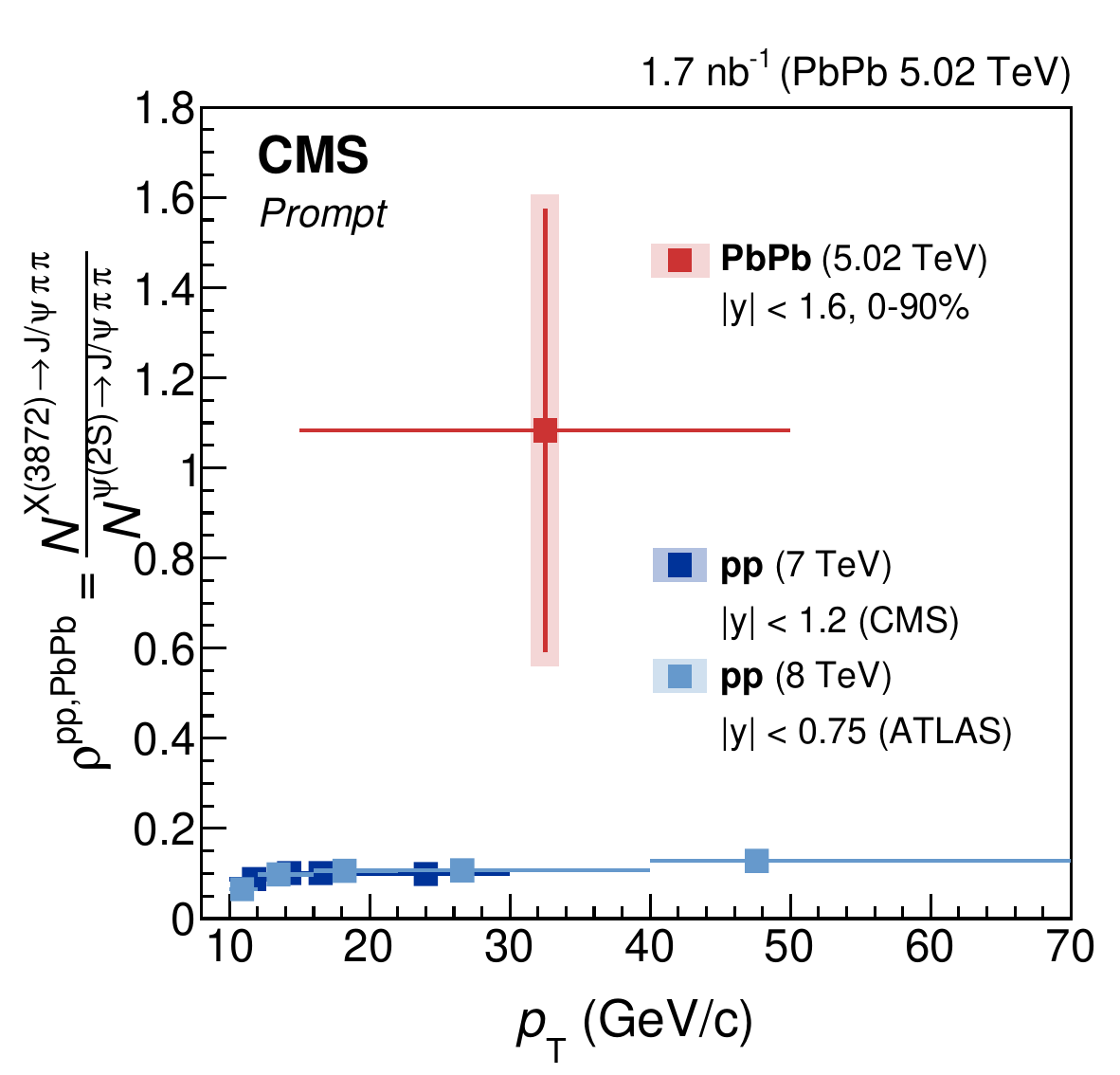}
\caption{The yield ratio $\rho^{\PbPb}$ of prompt \xthree over $\psi(2S)$ production in \PbPb collisions at $\sqrtsNN = 5.02\TeV$. 
The vertical bars (boxes) correspond to statistical (systematic) uncertainties.
The yield ratios $\rho^{\pp}$ of prompt \xthree over $\psi(2S)$ production in \pp collisions at $\sqrt{s} = 8\TeV$, measured by ATLAS~\cite{Aaboud:2016vzw}, and at $\sqrt{s} = 7\TeV$, measured by CMS~\cite{Chatrchyan:2013cld} are also shown.}
\label{fig:result}
\end{figure}

The ratio $\rho^{\PbPb}$ between the prompt \xthree and \psitwoS mesons is shown in \figref{fig:result}, together with $\rho^{\pp}$ measured as a function of \pt. The \pp data were measured at $\sqrts=7$ and 8\TeV, in the $\abs{y}<1.2$ and $\abs{y}<0.75$ intervals, respectively~\cite{Chatrchyan:2013cld,Aaboud:2016vzw,Aaij:2013zoa}. The 7\TeV result was derived using the CMS Collaboration published ratio of the inclusive yields~\cite{Chatrchyan:2013cld} and prompt fractions~\cite{Chatrchyan:2011kc,Chatrchyan:2013cld}. From Fig.~\ref{fig:result} it is clear that the prompt $\rho^{\pp}$ does not depend significantly on collision energy or rapidity. In \pp collisions at $\sqrts = 8\TeV$, in the kinematic range of $16<\pt<22\GeVc$, the $\rho^{\pp}$ measured by the ATLAS Collaboration is $0.106\pm 0.008\stat\pm 0.004\syst$~\cite{Aaboud:2016vzw}. This is to be compared to the prompt $\rho^{\PbPb}$ measured in this Letter, $\rho^{\PbPb} = 1.08\pm0.49\stat\pm 0.52\syst$. 

In the interval $15<\pt<20\GeVc$, the yield of the prompt \psitwoS in \PbPb collisions was reported to be significantly suppressed with respect to \pp collisions, $\raa^{\psitwoS}=0.142\pm0.061\stat\pm0.020\syst$ ~\cite{Sirunyan:2017isk}. This leads, using \eqqref{Eq:raa}, to an $\raa^{\xthree}$ central value larger than 1 (\ie, enhancement of the prompt \xthree yield in \PbPb compared to \pp collisions). However, the uncertainties are such that $\raa^{\xthree}$ is compatible with 1 within $\sim$1 standard deviation, and with $\raa^{\psitwoS}$ within $\sim$2 standard deviations. Thus, it is possible that in \PbPb collisions, the prompt \xthree yield has either no suppression with respect to \pp collisions, or as much suppression as the \psitwoS state. The much larger data sample expected in Run 3 at the LHC will answer whether $\raa^{\xthree}$ is different from $\raa^{\psitwoS}$ and significantly above unity. It may answer whether the \psitwoS meson production (a bound state of a \PQc and $\PAQc$ quarks, with $r\sim0.9\unit{fm}$)~\cite{Kluberg:2009wc}, is affected differently by the medium produced in \PbPb collisions than the \xthree state (that could be made of \PQc, $\PAQc$, \PQu, and $\PAQu$ quarks, with a radius of $r\sim0.3\unit{fm}$ or $r>1.5\unit{fm}$), the difference in both size and quark content playing a role into their production mechanisms. It will also answer whether the  \xthree prompt state production is different in \PbPb collisions compared to \pp collisions. The question whether \xthree is a tetraquark or a molecule cannot yet be answered, because of the statistical limitation of the data, and the disagreement among models. For example, while the AMPT transport model~\cite{Zhang:2020dwn} predicts $\raa^\text{molecule}>>\raa^\text{tetraquark}$ with $\raa^\text{molecule}>1$, the TAMU transport model~\cite{Wu:2020zbx} predicts $\raa^\text{molecule}\sim\raa^\text{tetraquark}/2$ (albeit, considering only the \xthree from regeneration processes).

In summary, the first evidence for \xthree production in heavy ion collisions is presented using lead-lead collisions at a center-of-mass energy of $\sqrtsNN = 5.02\TeV$ per nucleon pair, recorded with the CMS detector. The \xthree state is reconstructed using the decay chain $\xthree\to \Jpsi\,\Pgpp\Pgpm\to \mu^+\mu^-\Pgpp\Pgpm$. The measurement is performed for transverse momentum values of the \xthree of $15<\pt < 50\GeVc$ and rapidity $\abs{y}<1.6$. The significance of the inclusive \xthree signal is 4.2 standard deviations. The ratio $\rho^{\PbPb}$ between the prompt \xthree and \psitwoS yields times their branching fractions into $\Jpsi\,\Pgpp\Pgpm$ is found to be $1.08\pm0.49\stat\pm 0.52\syst$, to be compared with typical values of 0.1 for \pp collisions. This result provides a unique experimental input to theoretical models of the \xthree production mechanism, and of the nature of this exotic state.

\begin{acknowledgments}

  We congratulate our colleagues in the CERN accelerator departments for the excellent performance of the LHC and thank the technical and administrative staffs at CERN and at other CMS institutes for their contributions to the success of the CMS effort. In addition, we gratefully acknowledge the computing centers and personnel of the Worldwide LHC Computing Grid and other centers for delivering so effectively the computing infrastructure essential to our analyses. Finally, we acknowledge the enduring support for the construction and operation of the LHC, the CMS detector, and the supporting computing infrastructure provided by the following funding agencies: BMBWF and FWF (Austria); FNRS and FWO (Belgium); CNPq, CAPES, FAPERJ, FAPERGS, and FAPESP (Brazil); MES (Bulgaria); CERN; CAS, MoST, and NSFC (China); COLCIENCIAS (Colombia); MSES and CSF (Croatia); RIF (Cyprus); SENESCYT (Ecuador); MoER, ERC PUT and ERDF (Estonia); Academy of Finland, MEC, and HIP (Finland); CEA and CNRS/IN2P3 (France); BMBF, DFG, and HGF (Germany); GSRT (Greece); NKFIA (Hungary); DAE and DST (India); IPM (Iran); SFI (Ireland); INFN (Italy); MSIP and NRF (Republic of Korea); MES (Latvia); LAS (Lithuania); MOE and UM (Malaysia); BUAP, CINVESTAV, CONACYT, LNS, SEP, and UASLP-FAI (Mexico); MOS (Montenegro); MBIE (New Zealand); PAEC (Pakistan); MSHE and NSC (Poland); FCT (Portugal); JINR (Dubna); MON, RosAtom, RAS, RFBR, and NRC KI (Russia); MESTD (Serbia); SEIDI, CPAN, PCTI, and FEDER (Spain); MOSTR (Sri Lanka); Swiss Funding Agencies (Switzerland); MST (Taipei); ThEPCenter, IPST, STAR, and NSTDA (Thailand); TUBITAK and TAEK (Turkey); NASU (Ukraine); STFC (United Kingdom); DOE and NSF (USA).

\hyphenation{Rachada-pisek} Individuals have received support from the Marie-Curie program and the European Research Council and Horizon 2020 Grant, contract Nos.\ 675440, 724704, 752730, and 765710 (European Union); the Leventis Foundation; the Alfred P.\ Sloan Foundation; the Alexander von Humboldt Foundation; the Belgian Federal Science Policy Office; the Fonds pour la Formation \`a la Recherche dans l'Industrie et dans l'Agriculture (FRIA-Belgium); the Agentschap voor Innovatie door Wetenschap en Technologie (IWT-Belgium); the F.R.S.-FNRS and FWO (Belgium) under the ``Excellence of Science -- EOS" -- be.h project n.\ 30820817; the Beijing Municipal Science \& Technology Commission, No. Z191100007219010; the Ministry of Education, Youth and Sports (MEYS) of the Czech Republic; the Deutsche Forschungsgemeinschaft (DFG), under Germany's Excellence Strategy -- EXC 2121 ``Quantum Universe" -- 390833306, and under project number 400140256 - GRK2497; the Lend\"ulet (``Momentum") Program and the J\'anos Bolyai Research Scholarship of the Hungarian Academy of Sciences, the New National Excellence Program \'UNKP, the NKFIA research grants 123842, 123959, 124845, 124850, 125105, 128713, 128786, and 129058 (Hungary); the Council of Science and Industrial Research, India; the Ministry of Science and Higher Education and the National Science Center, contracts Opus 2014/15/B/ST2/03998 and 2015/19/B/ST2/02861 (Poland); the National Priorities Research Program by Qatar National Research Fund; the Ministry of Science and Higher Education, project no. 0723-2020-0041 (Russia); the Programa Estatal de Fomento de la Investigaci{\'o}n Cient{\'i}fica y T{\'e}cnica de Excelencia Mar\'{\i}a de Maeztu, grant MDM-2015-0509 and the Programa Severo Ochoa del Principado de Asturias; the Thalis and Aristeia programs cofinanced by EU-ESF and the Greek NSRF; the Rachadapisek Sompot Fund for Postdoctoral Fellowship, Chulalongkorn University and the Chulalongkorn Academic into Its 2nd Century Project Advancement Project (Thailand); the Kavli Foundation; the Nvidia Corporation; the SuperMicro Corporation; the Welch Foundation, contract C-1845; and the Weston Havens Foundation (USA).

\end{acknowledgments}

\bibliography{auto_generated}  
\cleardoublepage \appendix\section{The CMS Collaboration \label{app:collab}}\begin{sloppypar}\hyphenpenalty=5000\widowpenalty=500\clubpenalty=5000\input{HIN-19-005-authorlist.tex}\end{sloppypar}
\end{document}

%% file: HIN-19-005-authorlist.tex
\vskip\cmsinstskip
\textbf{Yerevan Physics Institute, Yerevan, Armenia}\\*[0pt]
A.M.~Sirunyan$^{\textrm{\dag}}$, A.~Tumasyan
\vskip\cmsinstskip
\textbf{Institut f\"{u}r Hochenergiephysik, Wien, Austria}\\*[0pt]
W.~Adam, F.~Ambrogi, T.~Bergauer, M.~Dragicevic, J.~Er\"{o}, A.~Escalante~Del~Valle, M.~Flechl, R.~Fr\"{u}hwirth\cmsAuthorMark{1}, M.~Jeitler\cmsAuthorMark{1}, N.~Krammer, I.~Kr\"{a}tschmer, D.~Liko, T.~Madlener, I.~Mikulec, N.~Rad, J.~Schieck\cmsAuthorMark{1}, R.~Sch\"{o}fbeck, M.~Spanring, W.~Waltenberger, C.-E.~Wulz\cmsAuthorMark{1}, M.~Zarucki
% ...
\vskip\cmsinstskip
\dag: Deceased\\
1:  Also at Vienna University of Technology, Vienna, Austria\\
% ...
